\documentclass{edm_article}
\usepackage{tabularx} 
\usepackage{booktabs} 
\usepackage{graphicx} 
\usepackage{array}

\begin{document}

\title{Predicting Cognitive Load Using Sensor Data in a Literacy Game}


%
%
%
%

\numberofauthors{2} 
\author{
%
%
\alignauthor
Minghao Cai\\
       \affaddr{EdTeKLA, Computing Science}\\
       \affaddr{University of Alberta}\\
       \affaddr{Edmonton, Canada}\\
       \email{minghaocai@ualberta.ca}
\alignauthor
Carrie Demmans Epp\\
       \affaddr{EdTeKLA, Computing Science}\\
       \affaddr{University of Alberta}\\
       \affaddr{Edmonton, Canada}\\
       \email{cdemmansepp@ualberta.ca}
}

\maketitle

\begin{abstract}
Educational games are being increasingly used to support self-paced learning. However, educators and system designers often face challenges in monitoring student affect and cognitive load. Existing assessments in game-based learning environments (GBLEs) tend to focus more on outcomes rather than processes, potentially overlooking key aspects of the learning journey that include learner affect and cognitive load. To address this issue, we collected data and trained a model to track learner cognitive load while they used an online literacy game for English. We collected affect-related physiological data and pupil data during gameplay to enable the development of models that  identify these latent characteristics of learner processes. Our model indicates the feasibility of using these data to track cognitive load in GBLEs. Our multimodal model distinguished different levels of cognitive load, achieving the highest Kappa ($.417$) and accuracy ($70\%$). Our model reveals the importance of including affect-related features (i.e., EDA and heart rate) when predicting cognitive load and extends recent findings suggesting the benefit of using multiple channels when modeling latent aspects of learner processes. Findings also suggest that cognitive load tracking could now be used to facilitate the creation of personalized learning experiences.

\end{abstract}


\section{Introduction}

Prior studies have found that educational games can facilitate self-paced learning \cite{acquah_digital_2020, stiller_game-based_2019, xu_scoping_2020}.
Despite the recognized benefits of game-based learning environments (GBLEs), research examining how learners react when interacting with these environments is still limited \cite{plass_handbook_2020}. Current evaluations of GBLEs often prioritize learning outcomes over the processes that lead to these outcomes \cite{petri_how_2017, xu_scoping_2020}. This approach may overlook elements of the learning experience, which limits our understanding of how learners react to these environments. Cognitive load is a critical aspect of learning experience that has been shown to have a complicated relationship with learner performance \cite{paas_cognitive_2014}. According to Cognitive Load Theory (CLT), cognitive load refers to the demands placed on working memory by learning tasks and activities \cite{sweller_cognitive_2011, paas_cognitive_2003, paas_cognitive_2014}. 

A key challenge in capturing the full spectrum of learner cognitive load in GBLEs is the reliance on traditional measures such as self-report questionnaires. These methods only provide retrospective assessments after learning or sporadic snapshots of learner experience \cite{krell_editorial_2022}. In addition, they depend on learners' ability to self-assess and express their internal states, which can be challenging \cite{leppink_development_2013, ayres_using_2006}. These limitations underscore the need for continuous methods of tracking cognitive load so we can gain a deeper understanding of self-paced processes within GBLEs.

Some work has attempted to use brain-activity data to track variations in cognitive load, with electroencephalography (EEG) emerging as a prominent tool \cite{antonenko_using_2010}. However, EEG relies on obtrusive equipment; the EEG cap that must be worn might distract users from the task, potentially causing unwanted cognitive load. 

More recently, pupil diameter has emerged as an indicator of cognitive load \cite{marshall_index_2002,duchowski_index_2018}. This work built on foundational research that linked pupil dilation to increased problem difficulty \cite{hess_pupil_1964}, which is related to intrinsic load. 
Although using pupil diameter has been effective \cite{marshall_index_2002,duchowski_index_2018}, it is sensitive to eye movements and several extraneous factors (e.g., ambient lighting conditions and the eye-tracker's camera angle) \cite{beatty_pupillary_2000}. 

Up until this point, the use of sensors for estimating cognitive load has relied on a single channel of data, like pupil diameter. 
Given that learning activities are often complex and involve the coordination of cognitive processes and affective states, relying solely on information from a single channel limits our understanding of how and when students learn best. 
Moving beyond existing models that primarily focus on unimodal data, our study explored a multimodal approach to enhance predictive performance.

Prior studies suggest that affect and cognitive load appear to interact during learning \cite{plass_emotional_2016, fraser_cognitive_2015, kalyuga_rethinking_2016, cai_toward_2024}. Yet, there has not been a consensus on how to effectively integrate affect into cognitive-load modeling. Following from this, we explored the effectiveness of incorporating physiological signals indicative of affective responses (i.e., electrodermal activity [EDA] and heart rate) with pupil diameter to predict cognitive load. 
This multi-modal model of cognitive load demonstrates the feasibility of using non-invasive sensor data to track cognitive load in GBLEs without interrupting the learner, thus enabling a dynamic evaluation of the learning process.
Moreover, investigation of what the model learned revealed that features related to affect support the evaluation of cognitive load.

\section{Methods}

We conducted a case study, where participants interacted with an English literacy game. This study was approved by our institutional research ethics board. Each participant consented to the study and was compensated with a \$30 gift card. All participants interacted with the GBLE for roughly 60 minutes. None had previously used the GBLE. 

Data were collected from 35 English-language learners (ELLs), but we only used data from 34 due to considerable sensor noise in one participant's data. ELLs were aged 17 through 33 years ($M = 24.1$). The languages participants reported speaking at home included Mandarin ($n = 24$), Japanese ($n = 3$), and Persian (Farsi; $n = 3$). Gujarati, Korean, Spanish, and French were each spoken at home by one ELL.

\subsection{Game-based Learning Environment}

The employed GBLE, Dreamscape, was designed to provide adaptive learning experiences \cite{cai_toward_2024}. It blends reading-comprehension activities into a base-building game. Players engage in a fantasy world where they are tasked with building and safeguarding their virtual homeland against invading reveries (creatures akin to soldiers in traditional base-building games). 
In this system, learner progress in the game is directly linked to their learning-activity performance. Learners are required to correctly answer questions to facilitate their progress in the game. Incorrect answers do not result in penalties.

\subsection{Data Collection}

We collected both sensor and self-report data. The self-report data were used as labels for model training and evaluation. 

The sensing system used two non-invasive wearable sensors. Pupil Core, a glasses-shaped eye tracker, recorded dual eye movements at 200 Hz and provided gaze tracking and pupil dynamics. A wireless wristband (Empatica E4) was used to provide continuous tracking of EDA data at 4Hz and heart rate using 10-second spans. 

The self-report instrument captured participants' subjective experiences of cognitive load. Following from cognitive load theory \cite{salomon_television_1984}, a 10-item Likert scale questionnaire (1 - Strongly Disagree and 10 - Strongly Agree) was developed by adapting existing scales from other educational domains where Cronbach's $\alpha$ was greater than $.80$.
This instrument assesses the three sub-types of load (intrinsic with 3 items, extraneous with 2 items, and germane load with 2 items) in addition to cognitive load (1 item). We averaged the sub-type scores and then combined them with the cognitive load item score to compute the final cognitive load score. The final scores were then discretized into three levels: low, moderate, and high.

Self-report data were collected after a participant completed a task in the system (e.g., finished a game-play activity, submitted an answer) to ensure minimal interference with learning processes. The instrument was administered approximately every 5-minutes.

\section{Model Development}

\subsection{Data Cleaning and Preprocessing}

Sensor data is noisy and requires pre-processing to enable its use. We report data cleaning by sensor type.

For each participant, the pupil signal includes pupil diameters along with time stamps and an estimated confidence level for each recorded diameter. The confidence ranges from 0.0 (could not be detected) to 1.0 (very high certainty). 

Because blinks can introduce substantial noise when measuring pupil size, we first removed data from 250 milliseconds before the onset and 250 milliseconds after the moment of a blink, as detected by the device. 
To further reduce the risk of incorporating erroneous or flawed data caused by the device itself, we implemented a confidence filtering step. Data points with a confidence level below $.65$ were excluded. 
For data continuity, missing data due to removal were linearly interpolated. To further reduce noise, we applied a $3^{rd}$-order Butterworth filter with a 4 Hz cutoff to minimize high-frequency disturbances while retaining signal fidelity.

Given the sensitivity of wristband sensors to various types of noise, we applied a simple moving average (SMA) filter after exporting participants' EDA and heart rate signal records.

After data cleaning, we aligned the sensor data with the timestamped participant self-report data. Data recorded from when participants were completing the self-report form were excluded. This step resulted in a dataset that contained 312 segments.

\subsection{Feature Extraction}
Following data cleaning and preparation, our research proceeded to feature extraction from the segmented physiological data.

Recognizing the lack of established evidence for the choice of time window for cognitive load prediction during literacy activities, we experimented with different window durations for extracting physiological features. This empirical approach allowed us to ascertain an appropriate window for cognitive load prediction. We tested the predictive performance of physiological data from 6 windows (30s, 60s, 90s, 120s, 150s, and 180s) and selected the window with the highest Kappa.

We extracted 7 features from the pupil diameter data for each segment: 
the average size of the pupil diameter (AvgPD);
the maximum pupil size in the dataset  (MaxPD);
the minimum pupil size in the dataset (MinPD);
the average speed at which the pupil diameter changes (AvgPV);
the maximum speed at which the pupil diameter changes (MaxPV);
the largest continuous change in pupil diameter without direction change (MaxPC); and
 the frequency of changes in pupil diameter (PCF).

We extracted 9 features from the EDA data for each segment: 
the average value of the EDA (AvgE);
the standard deviation of the EDA (SDGE);
the maximum value of the EDA (MaxE);
the minimum value of the EDA (MinE);
the difference between the maximum and minimum EDA (RngE);
the average speed of change in EDA levels (AvgEV);
the maximum speed at which EDA changes (MaxEV);
the maximum continuous change in EDA without direction change (MaxEC); and
the frequency of changes in EDA (ECF).

We extracted 9 features from the heart rate data for each segment: 
the average value of the heart rate (AvgH);
the standard deviation of the heart rate (SDH);
the maximum value of the heart rate (MaxH);
the minimum value of the heart rate (MinH);
the difference between the maximum and minimum heart rate (RngH);
the the average speed of change in heart rate (AvgHV);
the maximum speed at which the heart rate changes (MaxHV);
the maximum continuous change in heart rate without direction change (MaxHC); and
the frequency of changes in heart rate (HCF).

\subsection{Classification Models}
To achieve our research goal, we implemented and examined six conventional machine learning classifiers: Naive Bayes (NB), Decision Tree (DT), Support vector machine with linear kernel (Linear SVM), Support vector machine with radial basis function kernel (RBF SVM), logistic regression (LR), random forest (RF). 

In addition to the above classifier types, two approaches to modelling were compared: a baseline unimodal approach and our multimodal approach. 
The unimodal baseline used Index of Pupillary Activity (IPA) \cite{duchowski_index_2018}, specifically measuring fluctuations in pupil diameter over time to indicate cognitive load.
The multimodal approach augmented the baseline model by including the EDA and heart-rate features as inputs. This combination of classifiers and input features resulted in the training and evaluation of 12 models.

\subsection{Model Training}
We used Scikit-Learn (1.3.2) to develop our models. Model training was conducted on a desktop PC (i7-9700 CPU, 64GB RAM).

The data was split, with 80\% allocated to training and 20\% to testing.

We performed hyperparameter tuning on these models by applying 4-fold cross-validation to the training data. We used grid search to identify the optimal hyperparameters for each combination of classifier type and input features. Kappa served as the evaluation metric during the hyperparameter tuning process. The hyperparameter search space can be found in Appendix \ref{appendix:hp}. We report the selected hyperparameters below. 

For NB, we used alpha = 1 with Laplace smoothing. DT was configured using `gini'. The Linear SVM and RBF SVM had a regularization parameter of 1.0 (i.e., C = 1.0) and used the `linear' and `rbf' kernels, respectively. For LR, the `lbfgs' solver was used with an `l2' penalty and C = 1.0. Finally, RF used 100 estimators. 

In addition to tuning the standard hyperparameters reported above, we tuned the time window for sensor signals. The best performing time windows are reported in the results section (see Table \ref{table:combined_model_performance_window}).

\subsection{Model Evaluation and Interpretation}
We used mlxtend (0.23.1) to conduct statistical tests to examine model performance.

We report Cohen's $\kappa$ \cite{cohen_coefficient_1960} alongside accuracy metrics due to the skewed distribution of class labels within our dataset. 
Cochran's $Q$ test was used to assess performance differences among models, and McNemar's test was used to identify pairwise differences. 

To explore what factors contributed to model predictions, we analyzed the Gini importance \cite{breiman_classification_2017} of features in our best-performing multimodal model.

\section{Results}

\subsection{Model Performance}
Table \ref{table:combined_model_performance_window} reports the peak performance metrics for both the multimodal model and the unimodal model.

Cochran's Q test of multimodal model performance indicated a difference in classifier performance when using all three feature sets ($Q = 15.357$, $p = .009$, $\eta^2_{Q} = .061$). The subsequent McNemar's tests (Table \ref{table:classifier_comparison}) identified differences between RF and NB as well as RBF SVM and NB. These statistical testing results and the descriptive statistics of multimodal model performance indicate that random forest best supported the prediction of learner cognitive-load levels when pupil, heart rate, and EDA information were used as inputs.

In contrast to the multimodal models, no differences were found across the unimodal models ($Q = 8.391,$ $p = .136$, $\eta^2_{Q} = .034$).

\begin{table}[h]
\centering
\caption{Model Performance and window size}
\begin{tabular}{llccc}
\hline
\multicolumn{2}{l}{\textbf{Classifier}} & \textbf{$\kappa$} & \textbf{Accuracy} & \textbf{Window (s)} \\
\hline
\multicolumn{5}{l}{\textit{Multimodal models}} \\

& Naive Bayes & .121 & .417 & 150 \\
& Decision Tree & .256 & .583 & 180 \\
& Linear SVM & .210 & .517 & 60 \\
& RBF SVM & .319 & .633 & 210 \\
& Logistic Regression & .242 & .550 & 90 \\
& \textbf{Random Forest} & \textbf{.417} & \textbf{.700} & \textbf{210} \\

\multicolumn{5}{l}{\textit{Unimodal models}} \\

& Naive Bayes & .169 & .550 & 180 \\
& Decision Tree & .102 & .517 & 180 \\
& Linear SVM & .178 & .600 & 210 \\
& RBF SVM & .234 & .683 & 90 \\
& Logistic Regression & .159 & .517 & 90 \\
& Random Forest & .102 & .517 & 180 \\
\hline
\multicolumn{5}{l}{\footnotesize{Bold font indicates the best result}} \\
\end{tabular}
\label{table:combined_model_performance_window}
\end{table}

\begin{table*}[t]
\centering
\caption{McNemar's test results comparing multimodal classifiers}
\begin{tabular}{l|p{0.64in}p{0.64in}p{0.64in}p{0.64in}p{0.64in}}
\hline
 & \multicolumn{1}{c}{DT} & \multicolumn{1}{c}{Linear SVM} & \multicolumn{1}{c}{RBF SVM} & \multicolumn{1}{c}{LR} & \multicolumn{1}{c}{RF} \\ 
\hline
NB  & $\chi^2 = 3.125$, p $= .077$, $\eta^2 = .063$ & $\chi^2 = 1.636$, p $= .201$, $\eta^2 = .033$ & $\boldsymbol{\chi}^2$\textbf{= 6.760},\textbf{ p = .009}, $\boldsymbol{\eta}^2 = \textbf{ .135}$ & \textbf{$\chi^2 = 2.909$}, p $= .088$, $\eta^2 = .058$ & $\boldsymbol\chi^2 $\textbf{ = 8.257}, \textbf{p = .004}, $\boldsymbol{\eta}^2$ = \textbf{.165} \\
DT &  & $\chi^2 = 0.615$, p $= .433$, $\eta^2 = .012$  & $\chi^2 = 0.333$, p $= .564$, $\eta^2 = .007$  & $\chi^2 = 0.167$, p $= .683$, $\eta^2 = .003$  & $\chi^2 = 2.579$, p $= .108$, $\eta^2 = .052$ \\
Linear SVM  &  &  & $\chi^2 = 2.882$, p $= .090$, $\eta^2 = .058$ & $\chi^2 = 0.333$, p $= .564$, $\eta^2 = .007$  & $\chi^2 = 5.261$, \textbf{p $= .022$}, $\eta^2 = .105$ \\
RBF SVM &  &  &  & $\chi^2 = 1.316$, p $= .251$, $\eta^2 = .026$ & $\chi^2 = 1.600$, p $= .206$, $\eta^2 = .032$ \\
LR &  &  &  &  & $\chi^2 = 3.522$, p $= .061$, $\eta^2 = .070$ \\

\hline
\multicolumn{6}{l}{\footnotesize{Bold font indicates significant differences}} \\

\end{tabular}
\label{table:classifier_comparison}
\end{table*}

\begin{table*}[ht]
\centering
\caption{McNemar's test results comparing multimodal to unimodal models}
\begin{tabular}{l|llllll}
\hline
  & \multicolumn{1}{c}{NB} & \multicolumn{1}{c}{DT} & \multicolumn{1}{c}{Linear SVM} & \multicolumn{1}{c}{RBF SVM} & \multicolumn{1}{c}{LR} & \multicolumn{1}{c}{RF} \\ \hline
$\chi^2$  & 2.000   & 0.533    & 1.000   & 0.474    & 0.118    & \textbf{4.481}   \\
p   & .157    & .465    & .317    & .491    & .732    & \textbf{.034}    \\
$\eta^2$ & .040    & .011    & .020    & .009    & .002    & \textbf{.090}    \\ \bottomrule
\multicolumn{7}{l}{\footnotesize{Bold font indicates significant differences}} \\
\end{tabular}

\label{table:between comparasion}
\end{table*}

To determine whether the multimodal model outperformed the unimodal one, we compared model performance across inputs for each algorithm. The multimodal version of RF outperformed the unimodal one (Table \ref{table:between comparasion}), showing the benefit of including complementary sensor signals.

\subsection{Feature Importance}
We analyzed the multimodal random forest model because it had the highest $\kappa$. 
Our results showed all feature categories contributed to model predictions. We found that both the EDA and heart rate group of features played an important role in cognitive load prediction. We reviewed the categories and specific features in decreasing order and summarized the three most important features from each signal.

\subsubsection{Pupil Features}
Overall, the pupil feature group provided the most information to the model (36.61\%). The most important feature category was frequency of changes in pupil diameter (PCF), followed by the average speed at which the pupil diameter changed (AvgPV) and the average size of the pupil diameter (AvgPD).

\subsubsection{Heart Rate Features}
The heart rate feature group provided a similar amount of information to the model (36.24\%). The most important feature category was the maximum speed at which the heart rate changed (MaxHV), followed by the minimum (MinH) and average (AvgH) heart-rate values.

\subsubsection{EDA Features}
The EDA feature group provided less information to the model (27.15\%). In this group, the most important feature category was the minimum value (MinE), followed by the frequency of changes (ECF) and the average value (AvgE).

\section{Discussion}

Our research is driven by the primary goal of building a model to track learners' cognitive-load levels in a GBLE for literacy.
Our results demonstrate the feasibility of creating predictive models to classify levels of cognitive load in this context. Notably, models employing RBF SVM and RF emerged as top performers (Table \ref{table:combined_model_performance_window}). These models achieved Cohen’s $\kappa$ scores exceeding .319 and .417 and demonstrated prediction accuracies ranging from 63\% to 70\%. Moreover, our multimodal strategy generally yielded higher Cohen’s kappa scores across classifiers, with the RF model showing particularly superior performance.

Participants' physiological responses were recorded as they naturally occured without intentionally inducing affective or cognitive states. This approach provides a more authentic representation of how learners react to and process the game-based learning environment. This approach contrasts with studies where conditions are manipulated to elicit specific learner states \cite{scrimin_does_2015}, which, while valuable for controlled experiments, may not fully capture how learners behave under normal conditions.

Our multimodal model's superior performance over that of the baseline unimodal classifier's, highlights the added value of incorporating diverse physiological channels.

\subsection{Feature Contributions to Prediction}
Our results showed the feasibility of developing a predictive model and provided insight into the underlying factors influencing cognitive load. 

While all features played a role in the prediction process, pupil-based metrics had the highest importance in our prediction model with the frequency, speed of changes in pupil diameter, and average pupil size being the most critical features. These findings align with existing research that indicated pupil dilation and constriction are controlled by the autonomic nervous system and can be influenced by cognitive effort \cite{duchowski_index_2018}.

Beyond the frequently used pupil features, our analysis highlights the contributions (63.76\% in total) of EDA and heart rate features. Until now, these features have only been considered indicators of affective states \cite{hussain_affect_2011}. Our findings suggest these sensor streams may also be indicative of cognitive states. This finding aligns with recent research highlighting the interplay between learner affect and cognitive load. Studies have suggested that emotions play a crucial role in cognitive processes, with different emotional states potentially enhancing or hindering learners' ability to comprehend and retain information \cite{egidi_how_2009, mills_being_2019}. More recently, Cai et al. conducted a study demonstrating that affect valence can predict cognitive load in game-based learning environments \cite{cai_toward_2024}. Our modeling study underscores the importance of considering both cognitive and affective factors when designing and evaluating educational content, suggesting that an integrated approach to monitoring and responding to both affective states and cognitive load could enhance student modeling.

Among the heart rate features, rapid heart rate changes as well as the heart rate's mean and trough value helped predict learner cognitive load levels. 
These findings highlight the significance of incorporating heart rate, particularly its fluctuations, as an indicator of cognitive load.  
The identification of heart rate as an effective indicator advances our ability to detect changes in load, particularly with the  widespread availability of wearable sensor technology such as smartwatches.   

Similarly, the identification of EDA as another indicator advances our ability to detect cognitive load with EDA's minimum and average values as well as fluctuations in EDA supporting this goal. This finding highlights the nuanced ways in which physiological responses correlate with cognitive demand. The present work expands the association between increases in EDA and emotional arousal \cite{posada-quintero_innovations_2020} to cognitive load. The prominence of minimum EDA as an important feature suggests that moments of lower emotional arousal may offer insights into cognitive load states, possibly reflecting periods of reduced cognitive effort or enhanced understanding. 
The inclusion of EDA in student modeling goes beyond conventional cognitive assessments to support the diagnosis of cognitive load and learner affect that might then be used to adapt the system to enhance engagement and learning \cite{dmello_selective_2013}.

\subsection{Towards Adaptation based on Cognitive Load }
The best-performing model in our study supported the prediction of learner cognitive load in a literacy game context. These results illustrate the utility of physiological data for predicting cognitive load that could then be used to dynamically capture this aspect of learner experience so that we can better understand learner behaviors in ways that are not achievable through self-report measures. 

Our model tracks cognitive load using real-time physiological responses that can provide continuous monitoring of learner cognitive state without disrupting the flow of learning. This capability should enable the dynamic adaptation of support and instructional strategies to
enhance the effectiveness of GBLEs and improve learning experience \cite{paas_cognitive_2003, van_merrienboer_cognitive_2005}. 

Our work provides a feasible way to track learner cognitive load that can then be used to inform adaptation. These models enable the identification of moments when cognitive load becomes overtaxed, which could result from overly challenging tasks or poor system feedback. This is the point at which automated adaptation would be implemented. Adaptations could include adjusted learning activities, feedback that encourages learners to take a break, or suggested alternative learning strategies. The identification of lower cognitive demand periods could be used to inform adaptation in a manner that enhances subsequent engagement and information retention \cite{paas_cognitive_2014}.

\section{Conclusions}

Despite the growing adoption of GBLEs, the exploration of how learner cognitive load fluctuates within these complex interactive environments is still restricted. This limitation appears to result from a lack of methods for continuously monitoring such aspects of learner experience. 

In this study, we created a predictive model that uses sensor data to represent learner physiological responses as input to a model that estimates cognitive load. Our results demonstrate that a combination of data on pupil diameter, EDA, and heart rate offers a reasonable estimation for three cognitive load levels, highlighting the added value of using mulitmodal data. The sensor data used as input to the multimodal classifiers highlights the importance of affect-related features. These results also indicate that both cognitive and affective factors need to be accounted for in the design and assessment of educational materials. We suggest that an approach combining the monitoring of affective states and cognitive load could improve student modeling.

\section{Acknowledgments}
This work was supported in part by funding from the Social Sciences and Humanities Research Council of Canada and the Natural Sciences and Engineering Research Council of Canada (NSERC), [RGPIN-2018-03834].
%
\bibliographystyle{abbrv}
%

\appendix
\section{Hyperparameter Search Space} \label{appendix:hp}
For the decision tree, we tried tried `gini' and `entropy'. 

For Linear SVM, RBF SVM, and Logistic Regression (LR), we tuned C. The following values were explored: 0.1, 1, 10, and 100.

We tuned the number of trees in the random forest, exploring values in the range of 20 through 100 using a step size of 20.

\balancecolumns
\end{document}